\documentclass[11pt]{article}
\usepackage{pslatex,amsmath,amssymb,epsfig,singlecolumn}

\begin{document}

\title{The Lotus-Eater Attack}
\author{Ian A. Kash\\
Computer Science Dept.\\
Cornell University\\
kash@cs.cornell.edu\\
\And
Eric J. Friedman\\
School of Operations Research\\ and Information Engineering\\
Cornell University\\
ejf27@cornell.edu\\
\And
Joseph Y. Halpern\\
Computer Science Dept.\\
Cornell University\\
halpern@cs.cornell.edu\\
}

\pagenumbering{arabic}

\newtheorem{theorem}{Theorem}[section]
\newtheorem{corollary}{Corollary}[section]
\newtheorem{lemma}{Lemma}[section]
\newtheorem{proposition}{Proposition}[section]
\newtheorem{definition}{Definition}[section]
\newtheorem{non-theorem}{Non-theorem}[section]
\newtheorem{claim}{Claim}[section]
\newcommand{\thm}{\begin{theorem}}
\newcommand{\lem}{\begin{lemma}}
\newcommand{\pro}{\begin{proposition}}
\newcommand{\dfn}{\begin{definition} \rm}
\newcommand{\rem}{\begin{remark}}
\newcommand{\xam}{\begin{example}}
\newcommand{\cor}{\begin{corollary}}
\newcommand{\prf}{\begin{proof}}
\newcommand{\ethm}{\end{theorem}}
\newcommand{\elem}{\end{lemma}}
\newcommand{\epro}{\end{proposition}}
\newcommand{\edfn}{\bbox\end{definition}}
\newcommand{\erem}{\bbox\end{remark}}
\newcommand{\exam}{\bbox\end{example}}
\newcommand{\ecor}{\end{corollary}}
\newcommand{\eprf}{\end{proof}}
\newcommand{\beqn}{\begin{equation}}
\newcommand{\eeqn}{\end{equation}}
\newcommand{\wbox}{\mbox{$\sqcap$\llap{$\sqcup$}}}
\newcommand{\bbox}{\vrule height7pt width4pt depth1pt}
\newcommand{\sat}{\mathit{sat}}

\newtheorem{observation}{Observation}[section]
\newcommand{\obs}{\begin{observation}}
\newcommand{\eobs}{\end{observation}}

\newcommand{\commentout}[1]{}
\newcommand{\M}{{\cal M}}
\newcommand{\cS}{{\cal S}}
\newcommand{\IR}{\mbox{$I\!\!R$}}
\newenvironment{RETHM}[2]{\trivlist \item[\hskip 10pt\hskip\labelsep{\sc #1\hskip 5pt\relax\ref{#2}.}]\it}{\endtrivlist}
\newcommand{\rethm}[1]{\begin{RETHM}{Theorem}{#1}}
\newcommand{\repro}[1]{\begin{RETHM}{Proposition}{#1}}
\newcommand{\relem}[1]{\begin{RETHM}{Lemma}{#1}}
\newcommand{\recor}[1]{\begin{RETHM}{Corollary}{#1}}

\newcommand{\erethm}{\end{RETHM}}
\newcommand{\erepro}{\end{RETHM}}
\newcommand{\erelem}{\end{RETHM}}
\newcommand{\erecor}{\end{RETHM}}
\newcommand{\BR}{\mathit{BR}}

\newcommand{\br}{\mathit{br}}

\maketitle

\begin{quote}
\emph{
They started at once, and went about among the Lotus-eaters, who did
them no hurt, but gave them to eat of the lotus, which was so
delicious that those who ate of it left off caring about home, and did
not even want to go back and say what had happened to them, but were
for staying and munching lotus with the Lotus-eater without thinking
further of their return.}
\end{quote}
\begin{flushright}
The~Odyssey \cite{odyssey}
\vskip 10pt

\end{flushright}

\begin{abstract}
Many protocols for distributed and peer-to-peer systems have the
feature that nodes will stop providing service for others once they
have received a certain amount of service.  Examples include
BitTorent's unchoking policy, BAR Gossip's balanced exchanges, and
threshold strategies in scrip systems.
An attacker
can exploit this by providing service in a targeted way to prevent
chosen nodes from providing service.
While such attacks cannot be prevented, we discuss techniques
that can be
used to limit the damage they do.  These techniques presume that a
certain number of processes will follow the recommended protocol, even
if they could do better by ``gaming'' the system.
\end{abstract}

\section{Introduction}\label{sec:intro}

Many current distributed and peer-to-peer systems have the feature
that they are {\em satiable}; they have users that (by design) will
stop providing service to others if they are themselves receiving a
sufficient quantity of service.  In many cases this is the product
of ``tit-for-tat-like'' designs, which attempt to combat free riding
by denying service to those who are not providing it.  While this
approach provides an incentive for cooperation, it has the
unfortunate side effect that if there is no service for a peer to
provide, then he will generally receive reduced or no service.
Ironically, this opens the systems up to an attack that we call the
{\em lotus-eater attack}: the attacker does no direct harm to any
peer; instead he supplies the service to some peers, thus satiating
them.  Once those peers are satiated, they stop providing service to
others.  The peers not being satiated by the attacker then receive
reduced or no service.

A wide range of systems are satiable and thus potentially vulnerable
to this attack.
In {\em direct reciprocity} systems like BitTorrent \cite{cohen} and BAR
Gossip \cite{bargossip}, peers trade with the best partners they can
find (BitTorrent) or stop trading when there is nothing they want (BAR
Gossip).  An attacker can prevent a peer from serving others by
being a good trading partner and satisfying all of its requests.
In {\em indirect reciprocity systems}, such as reputation systems
\cite{guha04,eigentrust} and
scrip systems \cite{fileteller02,karma03}, peers need to perform
service for others often enough to maintain a good reputation or
supply of money.  If an attacker can ensure that a peer maintains a good
reputation or supply of money despite any requests the peer makes, then
that peer will no longer provide service for others.  Even systems not
designed to be tit-for-tat-like may be satiable.
For example, a node in a sensor network might shut down to save power
if it has received all the updates it needs.

All of these systems have users that will stop providing service in
response to this attack.  However, the exact way that the attack is
carried out and the overall impact of the attack on the system varies
significantly.
Consider the case of BAR Gossip.
In most gossip protocols, nodes randomly select partners to pass
updates on to so that the updates spread through the entire system.
However, this allows nodes to free ride by
receiving updates while not using their own bandwidth to pass them on
to anyone else.  BAR Gossip encourages rational nodes to provide
service by having nodes give away updates on an exchange basis.
The downside is that a node following the protocol will not
continue to provide service when there are no more updates for it to
receive.  If an attacker successfully distributes all of the updates
to a large percentage of the nodes in the system, then the majority of
interactions will result in no updates being exchanged.  For those
nodes being satiated by the attack, this is a wonderful outcome; they
are receiving perfect service.  However, those nodes that are not
receiving the updates from the attacker will have few
opportunities to get the updates they need.  Since the updates in the
intended application of BAR Gossip (for example, a streaming video
service) are time sensitive, this minority
of nodes will miss updates and may find the service unusable.  By
changing who is satiated over time, the attacker could even make the
service intermittently unusable for all nodes.

Another context where this attack can be effective is in
a scrip system.  In these systems users are paid for providing service
in scrip, a currency issued by the system.  They can then redeem this
scrip later in exchange for service.
An optimal strategy for a rational agent in such a system is to
choose a threshold and provide service only when he has less than
that threshold amount of scrip \cite{scrip07}.  If an attacker can
ensure that an agent has a large amount of money (either by giving
money away, or providing cheap service to him), the agent will stop
providing service.
By targeting a user or users who control
important or rare resources, the attacker could prevent all users
from receiving certain kinds of services.  This type of behavior
occurs regularly in the traditional economy when companies sign an
exclusive contract or put particular lawyers on retainer to deny
others access to them.

Despite the attack being possible in BitTorrent, it seems likely to do
significantly less damage.  In BitTorrent, peers (known as {\em
leechers})  cooperatively download a file.  Each leecher has $k$ other
{\em unchoked} peers
to whom he provides pieces of the file.
These unchoked peers are mainly
leechers that have recently provided it with the most
service, but some may be chosen randomly ({\em optimistic unchokes})
to try and find better peers.  It is quite possible to
ensure that, excluding these random choices,  all of his unchoked
peers are controlled by the attacker.  However, since most leechers
are downloading more than they upload, this is often actually a net
benefit to the torrent.  Even targeting users that are uploading
more than they download seems likely to only modestly impair the
progress of the torrent, especially since the attacker must
contribute significant bandwidth of his own to make sure he stays
unchoked.  The attacker could try and target leechers who have rare
pieces to artificially create a ``last pieces problem,'' but
BitTorrent's rarest first policy does a good job of resolving this
problem~\cite{legout06}.

These three examples are cases where, to varying extents, a
lotus-eater attack can impair the performance of a system.  In order
to understand how the attack works in general and why the
effectiveness varies, we state an informal theorem that characterizes
the conditions under which an attacker can cause
nodes in a system to stop providing service, and develop a simple model
of how this lack of service can be used to harm the system.
Using that model we examine design principles that make systems
resilient to lotus-eater attacks.  Two of these are traditional:
tolerating non-random failures and making satiation hard by the use of
coding or a scrip system.  The other two are newer principles that
have received relatively little study: making use of obedient nodes
and encouraging altruism.

The remainder of this paper is organized as follows. In
Section~\ref{sec:BAR} we examine in detail the effectiveness of a
lotus-eater attack on BAR Gossip as well as changes to the algorithm
that can make it more robust.
In Section~\ref{sec:model} we present a model that abstracts the general
structure of systems built on tit-for-tat style mechanisms and state
an informal theorem
that captures the essential nature of the attack and the
possible avenues for preventing it.
In Section~\ref{sec:prevent} we
examine design principles relevant to preventing the attack and some
of the subtleties involved in implementing them.  We conclude in
Section~\ref{sec:conclusion} with discussion of some open questions
raised by this attack.

\section{Attacking BAR Gossip} \label{sec:BAR}

In a BAR Gossip system, a broadcaster is releasing updates that nodes
need to collect within a certain period of time.  For example, in a
streaming video application, the updates are frames of the video that
need to be received in time to display.  Each round, the broadcaster
sends each of the updates for that round to a random subset of the
nodes.  Nodes then gossip the updates through two protocols, which
each node can initiate once with a pseudorandomly chosen partner
(nodes have no control over who their partner will be).  In a
\emph{balanced exchange}, nodes exchange as many updates as possible on a
one-for-one basis.  In an \emph{optimistic push}, the node initiating the
push sends a list of recently released updates it has to offer and a
list of updates expiring relatively soon it needs.  The other node can
then receive a limited number of the recent updates in exchange for
older updates or junk data.  The push is optimistic on the part of the
initiator because he hopes he
receives useful updates in return.
In particular, if a node has no missing older updates, he
has nothing to gain by initiating an optimistic push and a rational
node will not.  These protocols are described in greater detail
in~\cite{bargossip}.

To mount an attack on BAR gossip, the attacker divides the nodes into
two groups.  The first group is the {\em satiated nodes}, to whom the
attacker attempts to provide as many updates as possible.  The second
group is the {\em isolated nodes}, to whom the attacker provides no
service.  If the attacker provides enough updates to the satiated
nodes, they will make relatively few and small balanced exchanges
because most of their updates are provided by the attacker.  This also
means they will rarely be missing very old updates and so will rarely
initiate optimistic pushes.  Since isolated nodes receive no service
from attacking nodes and limited service from satiated nodes, they
have few opportunities to trade for each update.

\begin{figure}[htb]
\centering \epsfig{file=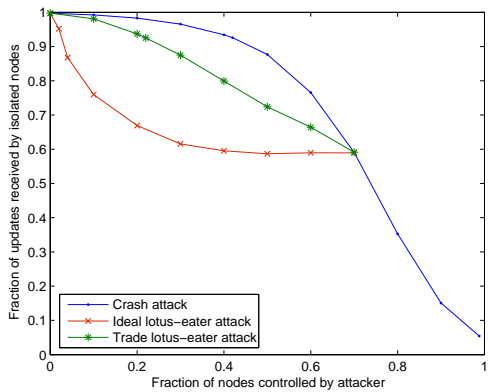, height=2.2in}
\caption{Three attacks on BAR Gossip.}
\label{fig:attack}
\end{figure}

In our experiments, the attacker attempts to satiate 70\%
of the system (including whatever percentage he controls).
The reason for our choice of 70\% will be explained shortly.
Figure~\ref{fig:attack} shows the results of three versions of the
attack on a BAR Gossip system using the same parameters as and an
updated version of the simulation from~\cite{bargossip}.  The
parameters are summarized in Table~\ref{table:params}.
In their simulation, nodes need to receive more than 93\%
of the updates for the stream to be usable.
The results in Figure~\ref{fig:attack} and later figures are given for
isolated nodes; satiated nodes receive near perfect service.

\begin{table}
\caption{Simulation Parameters}
\centering
\label{table:params}
\begin{tabular}{|l|c|}
\hline
Parameter & Value\\ [.5ex]
\hline
Number of Nodes & 250\\
Updates per Round & 10\\
Update Lifetime (rds) & 10\\
Copies Seeded & 12\\
Opt. Push Size (upd) & 2\\   [1ex]
\hline
\end{tabular}
\end{table}

The curve labeled ``crash attack'' provides a baseline where the
attacker
simply does nothing.  He may simply have crashed or be a Byzantine
node following the strategy of initiating but never completing
exchanges to waste bandwidth
(this was the strategy used by Byzantine nodes in~\cite{bargossip}).
With this attack, the attacker needs to control 42\% of the
system to ensure fewer than 93\% of the updates are delivered.
This curve is very similar to Figure~6
of~\cite{bargossip} where colluding nodes provide very little service
to others because they receive most updates from other colluding
nodes.  This curve also guided our decision to choose 70\%
as the fraction to satiate; it strikes a balance between the need to
satiate enough nodes to limit trade opportunities for isolated nodes
and a desire to isolate as many nodes as possible.

The ``ideal lotus-eater attack'' curve assumes that attacking nodes
can immediately send updates to all satiated nodes as soon as they
receive them.  This might be the case if the attacker can exploit the
implementation of the protocol to send updates to nodes with whom he
has not started an exchange.  Attacking nodes never trade, merely
forwarding all updates they receive from the broadcaster.  Note that
this means that satiated nodes will have to trade for any updates the
attacker did not receive from the broadcaster.  With this attack, the
attacker can control as few as 4\% of the system and still
make service unreliable.  Note that with so few nodes under his
control, the attacker is receiving only 39\% of the updates.
This shows that frequent partial satiation can be sufficient to attack
the system.

The ``trade lotus-eater attack'' curve makes the typically more
reasonable
assumption that the attacker can give updates to nodes only
during interactions dictated by the protocol.  However, he is able to
give nodes more updates than a normal node would (every update he
has).  Because the attacker needs to control enough nodes to
communicate with satiated nodes reasonably often, he needs
significantly more nodes than in the ideal lotus-eater attack; with
this version of the attack, the attacker needs to control
at least 22\% of the nodes in the system, far less than the
approximately 42\% percent that the more traditional
crash attack required.
This may make it possible to launch a lotus-eater attack in some
settings
where a crash attack would be impossible.
We should note, however, that this does require enough bandwidth at
each attacking node to satiate multiple nodes every round while the
crash attack requires essentially no bandwidth beyond that needed to
maintain the nodes in the system.

\begin{figure}[htb]
\centering \epsfig{file=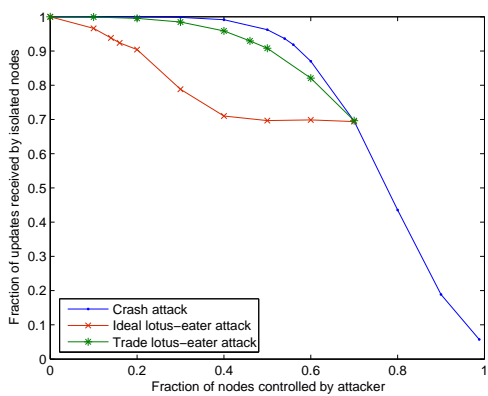, height=2.2in}
\caption{Larger push size reduces effectiveness.}
\label{fig:push}
\end{figure}

In anticipation of coming discussion, we also investigate the impact
of two changes on the effectiveness of lotus-eater attacks. First,
Figure~\ref{fig:push} shows the effect of increasing the maximum size
of an optimistic push to 10 updates.  This has the effect that nodes
that are willing to initiate optimistic pushes will be a more
altruistic towards other nodes; they are willing to give away more
updates at the risk of receiving junk.  This makes partial satiation
much less effective, so the ideal lotus-eater attack now requires at
least 15\% of the nodes in the system, which is enough to allow
him to provide 85\% of the updates to satiated nodes.  It also
makes the trade lotus-eater attack impractical, 
by nearly doubling the required fraction of nodes to 40\%.
This change does have two downsides.  First, rational nodes might no
longer be willing to participate in optimistic pushes if they tend to
receive significantly more junk updates due to the higher push size.
Second, Byzantine nodes can create more work by asking for a larger
number of updates.

\begin{figure}[htb]
\centering \epsfig{file=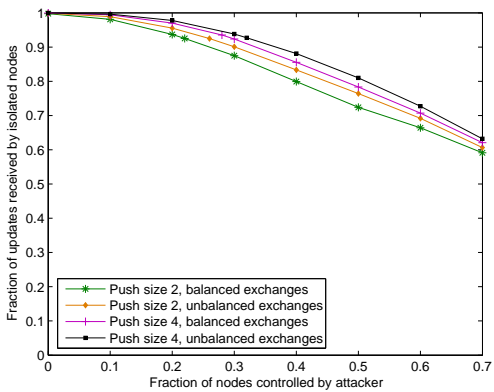, height=2.2in}
\caption{Obedient nodes reduce effectiveness.}
\label{fig:obedient}
\end{figure}

The other change we consider is to the behavior of the nodes in the
system.  In addition to Byzantine and rational nodes, the BAR model
includes the possibility of obedient ``altruist'' nodes who are
willing to follow the protocol even if it is not optimal.
One way we could exploit this is by allowing balanced exchanges to be
slightly unbalanced.  We modified the protocol so that nodes
are willing to give one more update than they receive, assuming they
are receiving at least one update.  Since the node already has the
overhead of a balanced exchange, it doesn't seem unreasonable that a
node would be willing to upload a little extra data.  Incentives to
free ride or exploit by Byzantine agents are minimal because there
must already be a balanced exchange occurring and there is only a
single additional update involved.  Figure~\ref{fig:obedient} shows
the effects of this change on a trade lotus-eater attack, both alone
and in conjunction with a more modest increase in the push size to 4.
The combination of these two small changes is enough to increase the
fraction of the system the attacker needs to control by almost 50\%.

\section{Understanding the Lotus-eater Attack}\label{sec:model}

In this section, we develop an understanding, at an abstract level,
of how lotus-eater attacks work and what can be done about them.
We first examine what properties of a system allow an attacker to use
a lotus-eater
attack, and
state a ``theorem'' that encapsulates the essence of the attack.
We then
examine how this attack, which is not directly harmful, can be used
to harm a system.  To do this, we develop a simple model whose
parameters characterize features of systems that
affect the viability of an attack.

A system is characterized by a graph $G = (V,E)$.  The nodes are the
users, each of which is a state machine implementing some protocol.
The edges are the pairs of nodes that can potentially communicate.
There is a set $T$ of labeled tokens;
one feature of a node's state is the set
of tokens that the node currently has.
A node may reach the point where it has all the tokens that
it wishes to collect.  This is captured by a {\em satiation function}
$s$, a monotone function that maps
a user $i$, a time $t$, and a set $T'\subseteq T$ of tokens  to $\{
\mathit{true, false} \}$.  Intuitively, $\sat(i,t,T') =
\mathit{true}$ if $i$ is does not need any more tokens at time $t$
if it has all the tokens in $T'$. A state $s$ for $i$ is
\emph{satiated} at time $t$ if $\sat(i,t,T') = \mathit{true}$, where
$T'$ is the tokens associated with $s$ at time $t$.

In a satiated state, a node has all of its current desires met.  It
may eventually leave the state if new tokens enter the system or it
loses some of its current tokens, but until that time it can gain no
benefit from other nodes.  Many protocols have the property that
a node in a satiated state will not provide service to other nodes.  In many
cases this design is due to a desire to make the protocol
incentive-compatible.  We adopt the term {\em
satiation-compatible} to describe protocols
where nodes in a satiated state do not provide service.%
\footnote{Note that incentive-compatible and satiation-compatible
are not equivalent notions.  It is easy to construct protocols that
satisfy one property and not the other.}
With these definitions in place we can state our ``theorem.''

\vskip 4pt

\obs \label{thm:attack}
In a system where a satiation-compatible protocol is used, an attacker
that can
provide a node with tokens sufficiently rapidly can prevent it from
ever providing service.
\eobs

\vskip 4pt

The observation is trivial.  In the extreme case, if an attacker
provides a full set of tokens instantaneously, then clearly the node
will  be satiated when a message arrives. The importance of this
observation is it helps abstract out the two key factors that allow
the lotus-eater attack to stop a node from providing service: a
satiation-compatible protocol and an attacker that provides tokens
``sufficiently rapidly.''  These two factors are what a protocol
designer must target in order to mitigate lotus-eater attacks.  In
Section \ref{sec:prevent}, we will analyze these two factors in more
detail.

While the observation tells us that an attacker can cause nodes to
stop providing service using a lotus-eater attack, it does not say
anything about why this would be a bad thing.
Indeed, for a (well-designed) system this
attack should have little or no negative impact.  In order to get a
sense of the ways this attack can actually damage a system, we
consider a simplified model of a token-collecting system.  This
system uses a simple protocol.  In each round, each
node $i$ selects up to $c$ communication partners from among
its
adjacent nodes and $i$ gets a copy of the tokens that each partner has,
while each partner gets a copy of the tokens $i$ has (for
simplicity, assume all of these events happen simultaneously).  Once
$i$ has a copy of all the tokens (i.e., once $i$ is satiated),
he stops communicating.
In many real systems, rather than stopping service
entirely, nodes actually continue to provide some service even through
they are satiated (for example seeding in BitTorrent).
We allow for this possibility in our model by having the probability
that a node responds to requests even when satiated be nonzero.
A system in this model is a tuple $(G,T,\sat,f,c,a)$ where:

\begin{itemize}

\item $G = (V,E)$ is the underlying graph, which we assume to be connected;

\item $T$ is a finite set of tokens;

\item $\sat(i,t,T') = \mathit{true}$ iff $T' = T$
(i.e. every node wishes to collect every token);

\item $f:V \rightarrow T$ is an initial allocation of tokens to nodes;

\item $c$ is a bound on the number of nodes that each node can contact
each round;

\item $a$ is the probability that a node responds to requests even if
satiated.  This captures the amount of altruism in the system.

\end{itemize}

We assume that, at the start of every round, an attacker chooses a
subset of the nodes and gives each node in the set all the tokens.
Clearly this overestimates the power of the attacker in most real
systems,
and ignores the possibility that $T$ will grow over time.
However, this simple model suffices to help us see where problems
may lie.

Of the six parameters in our model, $T$ and $\sat$ are
typically
beyond the control of the designer
(although,
as we discuss in Section \ref{sec:prevent}, techniques like
coding, which can be viewed as changing the set of tokens, may be of
some use in specific
cases).
Knowledge of
$G$, $f$, and $c$
can help an attacker know what to target; we discuss each of these three
parameters in turn, and then consider the role of $a$.

Suppose that the underlying graph $G$ is a grid.  Then at any time the
attacker can partition the graph with relatively little cost by
removing any
set of nodes that constitutes a cut.  If some side of the cut is
missing a token, nodes on that side of the cut will never be able to
collect all the tokens.  Clearly, an attacker can always make a cut
around a single node, but doing this on a large scale is expensive.
While finding inexpensive cuts depends on the structure of $G$, the
damage is significant only if some side of the cut is missing a token.
Whether this is the case depends (in part) on $f$, the initial allocation.
If many nodes start
with each token and those nodes are well spread, this attack is likely
to be ineffective.
(Note that, in a real system, what we are calling
the ``initial allocation'' may actually include some of the initial
exchanges, because an attacker cannot
always satiate instantly.)  This version of the attack is also
likely to be ineffective in random networks, but in, for example,
sensor networks, there is often an inherent structure an attacker
may be able to make use of.

Even in the absence of significant structure,
knowledge of the initial allocation may help an attacker,
particularly if one or more of the tokens
is  rare.  In the extreme case where some token is initially at a
single node, an attacker can deny the entire system access to that
token for the cost of satiating one node.
This version of the attack may be relevant
to networks set up for
file sharing, grid computing, and other similar applications, where some
resources is often rare.  Furthermore, in these systems it tends
to be relatively easy to determine what the rare resources and who has
them.

As an alternative to targeted removals, an attacker with sufficient
resources may simply attempt to satiate a large fraction of the system.
Here the parameter $c$ is relevant.
This parameter is,
in a sense, a measure of how many ``trade opportunities'' a
node gets each round.  If the attacker can successfully reduce the
number of trade opportunities, the overall rate at which tokens spread
through the system may decrease.  This approach is what we used in the
case of BAR Gossip, and it was particularly damaging there because the
updates had hard deadlines.

The final parameter $a$ is a factor that helps mitigate lotus-eater
attacks.  A system with $a > 0$ is not truly satiation-compatible
(and generally not truly incentive-compatible either, because agents
can often free ride on altruistically provided service).  However,
adding a little bit of altruism can make a big difference in
reducing the harm of attacks, since satiated nodes can still provide
some service.  For our simple model, any system with $a > 0$ will
eventually end up with all nodes satiated.
Although we capture the degree of altruism here by the probability
of responsiveness even when satiated, altruism can be introduced
into the system in other ways.  For example, seeding and optimistic
unchokes in BitTorrent and optimistic pushes in BAR Gossip
can all be viewed as ways to introduce some altruism into the
system. We discuss ways that altruism can be leveraged in greater
detail in Section~\ref{sec:prevent}.

\section{Preventing the Lotus-eater Attack}\label{sec:prevent}

Our observation says that if a satiation-compatible protocol is used
in a system, then a lotus-eater attack succeeds if an attacker can
provide service ``sufficiently rapidly.'' This suggests that one way
to prevent lotus-eater attacks is to abandon satiation-compatibility
entirely. In general this seems undesirable as many popular systems
are satiation compatible.  Furthermore, it seems difficult to design
a robust incentive-compatible system that is not also
satiation-compatible. Most designs for incentive-compatible systems
are tit-for-tat-like; they rely on some notion of reciprocity to
provide incentive-compatibility. Satiation-compatibility is a
natural consequence of this, because when a node is satiated there
is no room for reciprocity. Despite being among the simplest systems
in which to analyze the incentives of users, even BitTorrent is
still vulnerable to free riding \cite{JunAhamad05,largeview}.  So
abandoning these relatively simple systems to try and maintain
incentive-compatibility while avoiding satiation-compatibility seems
likely to introduce as many problems as it solves.
Since we do not wish to abandon satiation-compatibility entirely,
we focus on ways to tolerate lotus-eater attacks.
In this section, we examine four
design principles that can help do this: being resilient to
non-random failures, making satiation difficult, leveraging obedience, and
encouraging altruism.

Of the four principles, resilience to non-random failures
is the best studied; we have nothing new to add.
As we saw in Section \ref{sec:model}, attacks based on the
structure of $G$ and $f$ are
essentially independent of the fact that we are using a lotus-eater
attack.  These attacks work by removing key nodes from the system; the
way they are removed is essentially incidental.  A system vulnerable
to this type of attack is also vulnerable to many others, and may
experience difficulties even without an attack if key nodes happen to
become satiated.
We thus assume that $G$ and $f$ have been chosen to prevent this.

The second principle, making satiation hard, is more interesting.  As
a general principle, it is good even when an attack is not underway,
because less satiation means more opportunities for useful work to be
done.  In the context of our model, making satiation hard means
focusing on $T$ and $sat$.
While there may be an underlying set of tokens that a user wants to
collect, using a scrip system or reputation system effectively allows
the set of relevant tokens to be changed.
In such a system,
a node will determine satiation based on its current
amount of money or reputation.  This generally makes it easy
to satiate a few nodes, but difficult to satiate a large
number of nodes.  For example, in a scrip system there is generally a
fixed amount of money.  While it is easy for an attacker to
accumulate enough money to satiate a few nodes, there may not even be
enough money in the system to satiate a significant fraction
of the nodes.
This suggests that scrip could be the basis for an
incentive-compatible gossip system that is robust against lotus-eater
attacks.

In many systems, the goal is to collect a complete set of tokens.  A
node might need the complete set of updates or the all the pieces of a
file.  If a node only has a few tokens, he may be unable to trade with
most agents because they already have them making him ``effectively
satiated.''  Similarly, a node with almost every token may have a
hard time finding nodes with the remaining tokens he needs.
Another way to make satiation hard is to adopt policies that increase
the likelihood that nodes in such a situation will be able to make a
useful exchange.
BitTorrent uses a number of optimizations for these cases.  In general
it tries to avoid this effective satiation by using a ``rarest-first''
policy, where leechers will target rare pieces first.
When first joining the system, leechers will request random pieces to
get pieces to trade as quickly as possible.  Finally, BitTorrent has a
special ``endgame mode'' to allow for the rapid acquisition  of  the final
pieces~\cite{cohen}.
Another approach is to use  ideas from network coding, as done
by Avalanche~\cite{avalanche}, to change the requirements so that nodes
need to collect only enough independent tokens to reconstruct the full
information rather than the complete set of tokens.

The last two principles, leveraging obedience and encouraging altruism,
are perhaps the most interesting,
in terms of broad both applicability
and directions for future research.
Work in fault tolerance typically considers all nodes to be either
good or bad; work in game theory considers all nodes to be
rational.  But in practice, even in a system with rational nodes,
there will be a pool of users running the default client on the
default settings as long as this serves them reasonably well.  The BAR
model~\cite{barmodel} bridges this gap by
considering systems with a mix of Byzantine, rational, and
altruistic nodes. (We prefer to use the term \emph{obedient} for
what Aiyer et al.~call ``altruistic,'' since these are nodes that simply
follow the protocol, and use the term \emph{altruistic}
somewhat differently, using it to refer to nodes that provide service
even when satiated.  Of course, such nodes may be obedient as well, if
they are simply following the protocol.)
We know of only one protocol
that explicitly seeks to exploit these obedient
nodes~\cite{martin07}.
In the remainder of this section, we examine how obedience and
altruism can be used to prevent lotus-eater attacks.

One use of obedience is
to prevent sufficiently rapid satiation, by limiting
the rate at which an attacker can provide service.  Doing this
represents a radical departure from the typical design goal for most
P2P systems.   In general a designer strives to provide as much
service as rapidly as possible.  Now the goal becomes
that of providing service at a reasonable pace, and enforcing
that pace.
At first glance, reducing the rate at which a node provides service
seems like a silly idea.  However, there are a number of cases beyond
lotus-eater attacks in which this might be beneficial.  In general,
the incentive for a user to contribute to a system is that the
service he would receive if he free-rides is inferior.  If the
service provided to free-riders is good, there is little incentive
for participation. Limiting the amount of service provided can
increase the incentives
for cooperation and, in some cases, even make all nodes better off.  For
example, in a scrip system, if altruists are not handled appropriately
they can cause what would otherwise be a thriving economy to crash,
making all
agents worse off because they now receive only the level of service
altruists are providing~\cite{scrip07}.

The BAR Gossip
protocol \cite{bargossip} gives some insight into how the number of
nodes an attacking node contacts each round can be limited, but
limiting the amount of service the attacker provides in each trade is
more subtle.
Only two people know if an attacker provides excessive service: the
attacker and the node that benefits from it.
Suppose that we require a node to report if it is getting excessive
service from another node.
Since this excessive service is to its benefit,
a rational node might not report it.
But an obedient node would, if its protocol required it.
A node can use the signed messages generated by BAR Gossip to prove that
excessive service occurred, and get the reported node removed from they
system.  If
there are sufficiently many obedient nodes in the
system, then we can essentially prevent a lotus-eater attack.
Moreover, the cost of obedience will be low
if the attack is successfully prevented,
so it seems reasonable to
expect that a reasonable proportion of nodes will in fact be obedient.%
\footnote{We remark that this is an example
of the more general phenomenon that maintaining cooperation often
requires the existence of players
willing to incur costs \cite{Nowak07,kandori92}.}

Even if an attacker successfully satiates a large fraction of the
nodes, this will have no negative impact if the remaining nodes still
receive sufficient service.
One way to achieve this would be to increase the opportunities
that the remaining nodes have to trade.
The parameter $c$ describes the bound on
the number of peers a node has.  BitTorrent has caps on both the number
of open connections to maintain and the number of those connections to
unchoke.  BAR Gossip limits the number of exchanges per round to
minimize the damage Byzantine nodes can do.  However, these systems
need to make sure that $c$ is still large enough that the system
performs well.  Thus, selecting a good value for $c$ involves a
careful
balancing by the system designer.  We have seen that an attacker who
can satiate a large fraction of the nodes can effectively decrease $c$
to the point where performance becomes unacceptable.  This could be
prevented by increasing $c$ but, to guarantee the desired robustness,
$c$ might have to be unacceptably high.  An alternative is to increase
the value of $a$.  Adding enough altruism means that isolated nodes will
still receive service despite the attack.

One way to encourage altruism is to provide incentives for rational
nodes to behave in a way that ends up being altruistic.
This can be done by having nodes optimistically provide service in
exchange for the hope of return service.  If this generally ends up
being a net benefit for the node, a rational node will still
participate even though he might get away with providing less service.
In BitTorrent, even if every other leecher is satiated, a leecher will
still receive service through optimistic unchokes.
When a leecher in BitTorrent
optimistically unchokes another leecher, he is picking someone to send
data to in hopes of finding a reliable partner for the future.
We saw another example of this with the optimistic push protocol of
BAR Gossip in Figure~\ref{fig:push}.  Rational agents may be willing
to participate in large optimistic pushes if there is a reasonable
chance it will get them an update they would otherwise miss.

Another way to add altruism to a system is 
to leverage obedience by having
a protocol that requires nodes to behave altruistically.
In BitTorrent,
an isolated leecher gets service from seeds (who have already
downloaded the complete file).
Seeding in BitTorrent is not an incentive-compatible behavior.
Perhaps unsurprisingly, many leechers never remain to seed or seed
only
for a limited time  However, a sufficient number do seed to maintain
reasonably popular torrents.
In the context of BAR Gossip, Figure~\ref{fig:obedient} shows
that having nodes perform slightly unbalanced exchanges can make
lotus-eater attacks more difficult.

Increasing $a$ does make systems more robust, but there is
a tradeoff. The same mechanisms also provide a limited amount of
free service to all nodes, whether attacked or not.  If this amount
is too generous, nodes have an incentive to free ride using just
this amount of service. For example, in BitTorrent, a node that
connects to a large number of peers can get good service even if he
never uploads any data \cite{largeview}.
In many cases this incentive can be eliminated if nodes have to
perform nonproductive work in exchange for the altruistic service.
For example, in BAR Gossip, nodes that receive updates through an
optimistic push that have no updates to return must upload junk
data.

\section{Conclusion}\label{sec:conclusion}

The lotus-eater attack is, at least in the context of
incentive-compatible systems, an attack on the incentives of agents.
As incentive-compatible systems grow in popularity, we expect that
other ways will be found for an attacker to target systems through the
incentives of their users.  On a theoretical level, this points to the
need for a better understanding of equilibria in the presence of
Byzantine agents.  Some work has been done in this direction with
solution concepts like $k$ fault tolerant Nash
equilibria~\cite{eliaz02},  $(k,t)$-robust equilibria~\cite{ADGH06},
and BAR games~\cite{bargames}.  The last is the solution concept used
to analyze BAR Gossip.
What that definition in particular excludes
(at least with the assumption of risk-averse agents, which is typically
made in practice)
is the possibility
for Byzantine and rational nodes to collude either explicitly or, as
in the case of the lotus-eater attack, implicitly.
Thus, to the extent
that we want to provide guarantees about system performance in the
presence of both Byzantine and rational agents, we need a solution
concept that considers the possibility of such collusion.

Another concrete open problem that arises from this attack is how we
can design a system that limits the rate at which nodes can provide
service.  As we saw in Section \ref{sec:prevent}, this potentially is a
strong technique for preventing lotus-eater attacks by preventing an
attacker from providing service sufficiently rapidly to satiate
targeted nodes.  This problem seems to be relevant for other attacks
on incentives as well, since typically these require the attacker to
be ``too nice.''  Even if they are not explicitly attacking, nodes
that provide a disproportionate amount of service can become a point
of centralization in what is otherwise a decentralized system.

\section*{Acknowledgements}

We would like to thank Harry Li and Lorenzo Alvisi for allowing us to
use their BAR Gossip simulation.
EF, IK
and JH are supported in part by NSF grant ITR-0325453.  JH is
also supported in part by NSF grant
IIS-0534064
and by AFOSR grant 
FA9550-05-1-0055.

\bibliographystyle{abbrv}
\bibliography{Z:/Research/Bibliography/kash}

\end{document}